\documentclass[10pt, conference, compsocconf]{IEEEtran}
\ifCLASSINFOpdf
\else
 \usepackage[dvips]{graphicx}
 \graphicspath{{./figures/}}
 \DeclareGraphicsExtensions{.eps}
\fi
 \usepackage{algpseudocode}
 \usepackage{listings}
 \usepackage{xcolor}
\usepackage{multirow}

\begin{document}
%
\title{A New Method for Parallel Monte Carlo Tree Search}


\author{\IEEEauthorblockN{S. Ali Mirsoleimani\IEEEauthorrefmark{1}\IEEEauthorrefmark{2},
Aske Plaat\IEEEauthorrefmark{1},
Jaap van den Herik\IEEEauthorrefmark{1} and
Jos Vermaseren\IEEEauthorrefmark{2}}
\IEEEauthorblockA{\IEEEauthorrefmark{1}Leiden Centre of Data Science, Leiden University\\
 Niels Bohrweg 1, 2333 CA Leiden, The Netherlands}
\IEEEauthorblockA{\IEEEauthorrefmark{2}Nikhef Theory Group, Nikhef\\
 Science Park 105, 1098 XG Amsterdam, The Netherlands}}


%


\maketitle

\begin{abstract}
\textbf{
In recent years there has been much interest in the
Monte Carlo tree search algorithm, a new, adaptive, randomized optimization algorithm. In fields as diverse as Artificial Intelligence, Operations Research, and High Energy Physics,
research has established that Monte Carlo tree search can find good solutions without domain dependent heuristics. However, practice shows that reaching high performance on large parallel machines is not so successful as expected. This paper proposes a new method for parallel Monte Carlo tree search based on the pipeline computation pattern.
}
\end{abstract}


%
\IEEEpeerreviewmaketitle
\section{Abstract}

\section{Introduction}
A significant research effort has been put in the development of parallel algorithms for Monte Carlo Tree Search (MCTS) \cite{Chaslot2008,Schaefers2014,Mirsoleimani2015a}. Parallel algorithms for MCTS come in multiple versions \cite{Browne2012}. Among the available methods, \textit{tree parallelization} is one of the most suitable techniques for shared-memory platforms \cite{Chaslot2008}. Good scalability in tree parallelization has two distinct flavors \cite{Segal:2010:SPU:1950322.1950326,Mirsoleimani2015}: 

\begin{enumerate}

\item Playout scalability: Where the number of processors expands, and our goal is to minimize playing or computation time for MCTS. In this case, the scalability means speedup roughly proportional to the number of processors. For example, if a multiprocessor configuration with 16 processors is used and reduced the single-processor configuration time by a factor of close to 16, a perfect playout-speedup is achieved. 

\item Strength scalability: Where the number of processors expands, and our goal is to maintain a constant playing strength given a fix computational budget for larger degree of parallelism. For example, if a multiprocessor configuration with 16 processors is used and achieved the single-processor configuration playing strength given an identical computational budget, a perfect strength-speedup is achieved.  

\end{enumerate}

The scalability study of tree parallelization is suggesting that perfect strength-speedup beyond a certain number of cores may not be possible due to the scaling limits of single threaded search \cite{Segal:2010:SPU:1950322.1950326}. It is also shown that playing strength of MCTS will decrease to some degree with more parallel processors due to search overhead \cite{Kuipers2013a}. It is also shown that on many-core processors, perfect playout-speedup is really hard to achieve due to synchronization overhead \cite{Mirsoleimani2015a}.
 
Generally, implementing tree parallelization for MCTS often creates a dilemma. We can either use a small granularity algorithm with a high playout-speedup but also low strength-speedup, or we can use a larger granularity algorithm that has high strength-speedup but with a lower playout-speedup \cite{Mirsoleimani2015,Kuipers2013a}. A version of the parallel MCTS that is both playout-scalable and strength-scalable , has the best parallel efficiency. The purpose of this paper is to introduce an efficient method for parallel MCTS algorithm based on the \textit{pipeline} computation pattern that has both types of scalability.

The rest of this paper is organized as follows. In section \ref{sec:back} the
required background information is briefly discussed. Section \ref{sec:related} discusses related work. Section \ref{sec:mctspipe} describes the proposed method. Finally, in Section \ref{sec:conclusion} we conclude the paper.

\section{Background}
\label{sec:back}
\subsection{Monte Carlo Tree Search}
\begin{figure}
\centering
\includegraphics[scale=0.5]{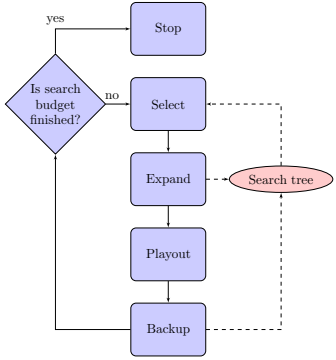}
\caption{The flowchart of the sequential MCTS algorithm.}
\label{fig:mcts-flow}
\end{figure}

Figure \ref{fig:mcts-flow} illustrates the flowchart of MCTS. The MCTS algorithm iteratively repeats four steps or operations to construct a search tree until a predefined computational budget $m$ (i.e., time or iteration constraint) is reached : 
\begin{enumerate}
\item \textit{Select}: Starting at the root node, a \textit{trajectory} that is a path of nodes inside the search tree is selected until a non-terminal-state node with unexpanded (i.e. unvisited) children is reached. Each of the nodes is selected based on a \textit{selection policy}. Among the proposed selection policies, the Upper Confidence Bounds for Trees (UCT) is one of the most commonly used policies~\cite{Kocsis2006,Browne2012}. A child node $j$ is selected to maximize: 
\begin{equation}
UCT=\overline{X}_{j}+C_{p}\sqrt{\frac{\ln(n)}{n_{j}}}
\end{equation}
where $\overline{X}_{j}=\frac{w_{j}}{n_{j}} $ is the average reward from child $j$, $w_{j}$ is the reward value
of child $j$, $n_{j}$ is the number of times child $j$ has
been visited, $n$ is the number of times the current node has been
visited, and $C_{p}\geq0$ is a constant. The first term in the UCT
equation is for \textit{exploitation} of known parts of the tree and the second term is for
\textit{exploration} of unknown parts~\cite{Browne2012}. The level of exploration of the UCT
algorithm can be adjusted by the $C_{p}$ constant.


\item \textit{Expand}: One of the children of the selected non-terminal-state node is generated and added to the selected trajectory.
\item \textit{Playout}: From the given state of the newly added node, a sequence of simulated actions is performed until a terminal state in the state space is reached.
\item \textit{Backup}: The terminal state is evaluated to produce a reward value $\Delta$. In the selected trajectory, each node's visit count $n$ is incremented by 1 and its reward value $w$ updated according to $\Delta$ \cite{Browne2012}.
\end{enumerate}
As soon as the computational budget $m$ is exhausted, a child of the root node is selected.

\subsection{Parallel MCTS}
The independent nature of each playout in MCTS means that the algorithm is a good target for parallelization. However, efficient parallel implementation of MCTS algorithm is challenging due to 3 types of overhead:
\begin{enumerate}
\item \textit{Communication overhead} refers to the cost of sending a
message from one processor to another over a network. 
\item \textit{Synchronization overhead} is the idle time that some
processors have to wait for the others to reach the
synchronization point.
\item \textit{Search overhead} occurs when a parallel implementation
of a search algorithm searches useless part of a
search space more that the sequential version.

\end{enumerate}
Different approaches for parallelizing MCTS are available but all of them are suffering from one or more of the overheads. These overheads are not independent. For example, reducing search overhead usually increases synchronization and communication overhead \cite{Soejima2010}. Furthermore, search overhead has an inverse relation with strength scalability and synchronization and communication overheads have an inverse relation with playout scalability. Therefore, the goal is to develop a method that is minimizing the search overhead without increasing the other two overheads. 
\subsection{Pipeline Computation Pattern}

A car assembly line is a good analogy for the pipeline computation pattern. The manufacturing process can be broken down into a sequence of operations each of which adds some component, say the engine or the windshield, to the car. An assembly line (pipeline) assigns a component to each worker. As each car moves down the assembly line, each worker installs the same component over and over on a succession of cars. After the pipeline is full (and until it starts to empty) the workers can all be busy simultaneously, all performing their operations on the cars that are currently at their stations.
The possibility of simultaneously performing different operations on different data elements is the potential concurrency which pipeline pattern exploit. Each task consists of repeatedly applying an operation to a data element, and the dependencies among tasks are ordering constraints enforcing the order in which operations must be performed on each data element.

\section{Related Work}
\label{sec:related}
Below we review related work on MCTS parallelizations.
The two major parallelization methods for MCTS are root parallelization and tree
parallelization~\cite{Chaslot2008}. There are also other techniques such as leaf
parallelization~\cite{Chaslot2008} and approaches based on transposition table driven
work scheduling (TDS)~\cite{Yoshizoe2011a}~\cite{Romein99}.
\begin{enumerate}

\item Tree parallelization:
For shared memory machines, tree parallelization is a suitable method. It is used in {\sc Fuego}, an open source Go program \cite{Enzenberger2010b}. In tree parallelization one MCTS tree is shared among several threads that are performing simultaneous searches~\cite{Chaslot2008}. It is shown in~\cite{Chaslot2008} that the playout speedup of tree parallelization with \textit{virtual loss} does not scale perfectly for up to 16 threads. The main challenge is the use of the data locks to prevent data corruption. Moreover, it is shown in ~\cite{Enzenberger2010a} that a lock-free implementation of this algorithm provides better scaling than a locked approach. In~\cite{Enzenberger2010a} such a lock-free tree parallelization for MCTS is proposed. They intentionally ignored rare faulty updates inside the tree and studied the scalability of the algorithm for up to 8 threads. Baudi\v{s} et al.~\cite{citeulike:10600156} reported the performance of a lock-free tree parallelization with virtual loss for a Go program. The strength-speedup in only self-play experiments is good up to 16 cores but the improvement drops after 22 cores. 

\item Root parallelization:
Chaslot et al.~\cite{Chaslot2008} reported results that root parallelization shows perfect playout speedup for up to 16 threads.
Soejima et al.~\cite{Soejima2010} analyzed the performance of root parallelization in detail. They showed that a Go player that uses lock-free tree parallelization with 4 to 8 threads outperformed the same program with root parallelization which utilizes 64 distributed CPU cores. This result suggests the superiority of tree parallelization over root parallelization in shared memory machines.
Fern and Lewis‍ thoroughly investigated an Ensemble UCT
approach in which multiple instances of UCT were run independently. 
Their root statistics were combined to yield the final result~\cite{fern2011ensemble}. 
\end{enumerate}

\section{The proposed method}
\label{sec:mctspipe}
In this section the potential concurrency in MCTS will be exploited. The first step is to decompose MCTS into tasks. The second step is to find out how these tasks must be ordered to satisfy dependencies among them. Finally, the pipeline pattern is used to introduce parallelism into MCTS.

\subsection{Defining the Tasks and Dependencies}
The first step towards designing a parallel algorithm is to find out concurrent tasks in MCTS. There are two levels of task decomposition in MCTS:
\begin{enumerate}

\item \textit{Iteration-level task (ILT)}: In MCTS the computation associated with each playout is a separate task. Therefore, it might work well to base a task decomposition on mapping each iteration onto a task. 
\item \textit{Operation-level task (OLT)}: Another level of task decomposition for MCTS is happening inside each iteration. Each of the four MCTS steps can be treated as a separate task. 

\end{enumerate}
In parallel MCTS, a shared search tree is modified by multiple tasks simultaneously and therefore serves as a source of dependency among them. In MCTS, there are two levels of data dependency:

\begin{enumerate}

\item \textit{Iteration-level dependency} (ILD): In MCTS, iteration \textit{n} requires updates from the previous iterations to select an optimal trajectory. However, it is possible to compromise over this requirement and conducting required computations for iteration \textit{n} independently from the iteration \textit{n-1}. Therefore, there is a \textit{soft dependency} between two consecutive iterations. However, This compromise produces search overhead that means exploring part of the tree which are not important or performing duplicate search. 

\item \textit{Operation-level dependency} (OLD): Each of the four steps in MCTS has a \textit{hard dependency} to its previous step. For example, the expansion of a trajectory cannot occur until the selection step is completed, the playout can not be performed until a trajectory being expanded and the backup step also needs the output of the playout step.

\end{enumerate}


\subsection{Designing the Pipeline}
There are two possible types of parallelism for MCTS based on ILT and OLT:
\begin{enumerate}
\item \textit{Iteration-level Parallelism} (ILP): The tasks in this method are from the type of ILT and they only have ILD dependency. 

\item \textit{Operation-level Parallelism }(OLP): The tasks in this method are from the type of OLT and they have both ILD and OLD.

\end{enumerate}
\begin{figure}
\centering
\includegraphics[scale=0.6]{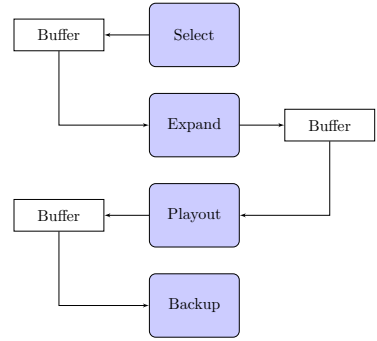}
\caption{Linear MCTS pipeline.}
\label{fig:mcts-pipe}
\end{figure}

The pipeline pattern can introduce parallelism into MCTS using OLP. Figure \ref{fig:mcts-pipe} shows a linear MCTS pipeline. The potential concurrency in MCTS is exploited by assigning each of the steps to a different stage of the pipeline and having them work on separate processing elements (PEs) simultaneously, with the trajectories passing from one stage to the next through buffers as operations are completed. This method is the most straightforward way to deal with ordering constraints among OLTs to satisfy requirements of OLD.

\begin{figure}
\centering
\includegraphics[scale=0.7]{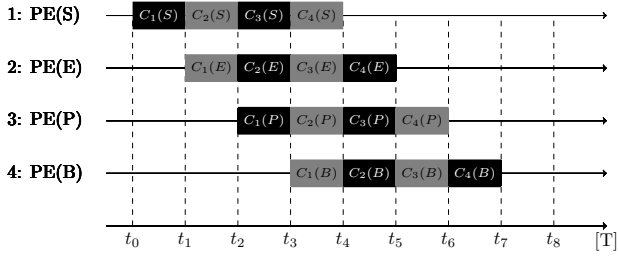}
\caption{Scheduling diagram for MCTS pipeline with equal stages.}
\label{fig:mcts-pipe-1}
\end{figure}

Figure \ref{fig:mcts-pipe-1} illustrates how the tasks in MCTS pipeline execute over time. Let $C_{i}$ represent a multiple-step computation on trajectory $i$. $C_{i}(j)$ is the jth step of the computation in MCTS (i.e., $j \in$ $\lbrace$S, E, P and B$\rbrace$). The idea is to map computation steps to pipeline stages so that each stage of the pipeline computes one step. Initially, the first stage of the pipeline performs $C_{1}(S)$. After that completes, the second stage of the pipeline receives the first trajectory and computes $C_{1}(E)$ while the first stage computes the first step of the second trajectory, $C_{2}(S)$. Next, the third stage computes $C_{1}(P)$, while the second stage computes $C_{2}(E)$ and the first stage $C_{3}(S)$. 

Assume each stage of the pipeline takes exactly the same amount of time to do its work, call that duration T. Figure \ref{fig:mcts-pipe-1} shows that the expected execution time for 4 trajectories in a MCTS pipeline with four stages is approximately $7\times T$. In contrast, the sequential version takes approximately $16\times T$ because each of the 4 trajectories must be processed one after another.

Notice that concurrency is initially limited and some PEs remain idle until all the stages are occupied with useful work. This is referred to as filling the pipeline. At the end of the computation (draining the pipeline), again there is limited concurrency and idle PEs as the final trajectory works its way through the pipeline. 



\subsection{MCTS Pipeline with Unequal Stages}

\begin{figure}
\centering
\includegraphics[scale=0.6]{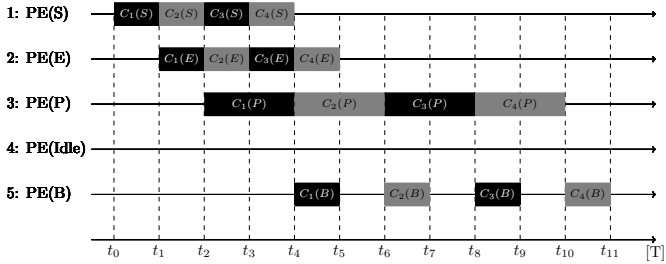}
\caption{Scheduling diagram for MCTS pipeline with unequal stages.}
\label{fig:mcts-pipe-sche-single}
\end{figure}

The pipeline pattern works better if the operations performed by the various stages of the pipeline are all about equally computationally intensive. If the stages in the pipeline vary widely in computational effort, the slowest stage creates a bottleneck for the aggregate throughput. In other words, when there are enough PEs for each pipeline stage, the speed of a pipeline is approximately equal to the speed of it's slowest stage. In MCTS, the stages are not equally computationally intensive, i.e., generally the playout step is more computationally expensive than other steps.
For example, Figure \ref{fig:mcts-pipe-sche-single} shows the scheduling diagram that occurs when the playout stage takes $2\times T$ units of time while others take T units of time.  
You can see that it's not possible to keep all the PEs completely busy (work execution in backup stage is fragmented). On average (with a large number of trajectories when the pipeline is full), the time to process a trajectory is $2\times T$. Figure \ref{fig:mcts-pipe-sche-single} shows the expected execution time for 4 trajectories is approximately $11\times T$. 

\begin{figure}
\centering
\includegraphics[scale=0.6]{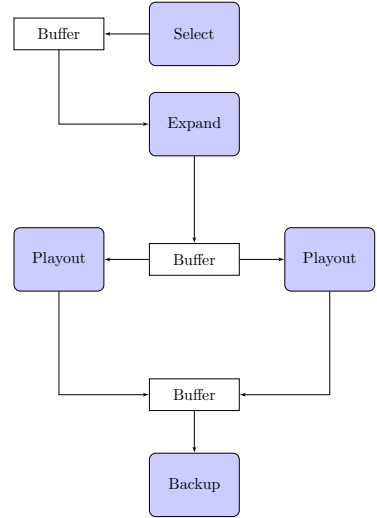}
\caption{Nonlinear MCTS pipeline.}
\label{fig:mcts-pipe-chart-multi}
\end{figure}

\begin{figure}
\centering
\includegraphics[scale=0.6]{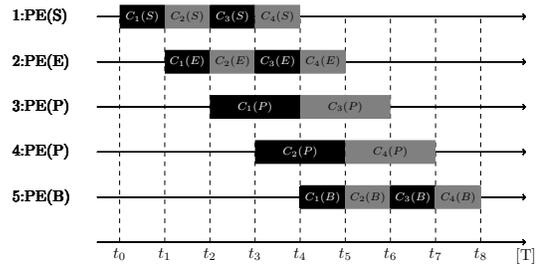}
\caption{Scheduling diagram for MCTS pipeline with load balancing.}
\label{fig:mcts-pipe-sche-multi}
\end{figure}

Figure \ref{fig:mcts-pipe-chart-multi} shows a nonlinear MCTS pipeline with two parallel playout stages. Both of the playout stages take trajectories produced by the expand stage of the pipeline. Using two playout stages to load balance the pipeline results in an overall speed of approximately $T$ units of time per trajectory as the number of trajectories grows. 
Figure \ref{fig:mcts-pipe-sche-multi} shows the MCTS pipeline is perfectly balanced by using two playout stages and the expected execution time for 4 trajectories is approximately $8\times T$.




Introducing parallel stages can make the MCTS pipeline more scalable. A parallel stage is different from a serial stage with internal parallelism, because parallel stage can process multiple input items at once and can deliver the output items out of order. The introduction of parallel stages introduces a complication to serial stages. In a pipeline with only serial stages, each stage receives items in the same order. But when a parallel stage intervenes between two serial stages, the later serial stage can receive items in a different order from the earlier stage.

\section{Conclusion}
In this paper, we proposed a new method to parallelize Monte Carlo Tree Search by utilizing pipeline pattern. The advantage of the proposed algorithm is that it could show both strength-speedup and playout-speedup. 
\label{sec:conclusion}

\section*{Acknowledgment}
This work is supported in part by the ERC Advanced Grant no. 320651, “HEPGAME.”

\IEEEtriggeratref{8}


\bibliographystyle{IEEEtran}
\bibliography{IEEEabrv,Bib-mctspipe}

\end{document}